**Pulsed Laser Deposition of LaAlO$_3$-Ba$_{0.5}$Sr$_{0.5}$TiO$_3$ Thin Films for Tunable Device Applications**

L. B. Kong[1], L. Yan[2], K. B. Chong[2], C. Y. Tan[2], L. F. Chen[1] and C. K. Ong[2]

[1]*Temasek Laboratories, National University of Singapore,*

*10 Kent Ridge Crescent, Singapore 119260*

[2]*Center for Superconductor and Magnetic Materials, National University of Singapore,*

*Lower Kent Ridge Crescent, Singapore 119260*

LaAlO$_3$-Ba$_{0.5}$Sr$_{0.5}$TiO$_3$ (LAO-BST) thin films, with different LAO contents, were deposited on Pt/Ti/SiO$_2$/Si and (100) LaAlO$_3$ substrate, via a pulsed laser deposition (PLD). Phase composition and microstructure of the LAO-BST thin films were characterized by X-ray diffraction (XRD) and scanning electron microscopy (SEM). The LAO-BST thin films were solid solution of LAO and BST, with lattice constant decreasing with increasing LAO content. The composition of (LAO)$_x$(BST)$_{1-x}$, estimated from the lattice constant of the thin films derived from the targets of 1/6, 2/6 and 3/6 LAO, was 0.143, 0.278, 0.476, respectively. The grain size of the thin films decreased as a result of the incorporation of LAO into BST. Dielectric properties of the LAO-BST thin films on Pt/Ti/SiO$_2$/Si were measured at low frequency (100 kHz), while those on LaAlO$_3$ substrate were characterized at high frequency (~7.9 GHz). The dielectric constants of the film derived from pure BST target and those from the targets with 1/6, 2/6 and 3/6 area ratio of LAO, on Pt/Ti/SiO$_2$/Si substrate, were 772, 514, 395 and 282, with a dielectric loss of 0.096, 0.023, 0.024 and 0.025, and a dielectric tunability of 65%, 53%, 43% and 14%, respectively. The dielectric constant of the LAO-BST films on LaAlO$_3$ substrate were 2436, 2148, 2018, 1805, with a maximum dielectric tunability of 11%, 13%, 10% and 8% at a maximum applied voltage of 2.4 kV (~9.2 kV/cm).

**I. Introduction**

High dielectric tunability and low dielectric loss are basic requirements for microwave device applications, such as tunable oscillators, phase shifters and varactors [1-3]. Ferroelectric materials, for example, barium strontium titanate ($Ba_{1-x}Sr_xTiO_3$ or BST), have been acknowledged to be the most promising candidates for these applications, due to their large electrical field dependent dielectric constant and fast responses [4-6]. Usually, ferroelectrics with a Curie temperature below operating temperature are used for practical device applications, since paraelectric state of ferroelectric materials has lower dielectric loss due to the disappearance of hysteresis. This is the reason why barium strontium titanate with composition around $Ba_{0.5}Sr_{0.5}TiO_3$ (BST) has been widely used for room temperature applications [7-10].

A critical issue for practical device application of ferroelectric materials is the reduction of the dielectric losses [11]. Many efforts have been made for this purpose in recent years. It has been found that doping of other oxides with low dielectric losses into ferroelectric materials is an effective way to reduce the dielectric losses. Sengupta and Sengupta [12, 13] were among those who tried to reduce dielectric loss of $Ba_{0.5}Sr_{0.5}TiO_3$ ceramic, thick and thin films, by the addition of MgO, $Al_2O_3$ and $ZrO_2$. They found that MgO was the best dopant in terms of reduction in dielectric losses. This, therefore, led to extensive studies on the doping effect of MgO on dielectric properties of $Ba_{1-x}Sr_xTiO_3$. For example, Joshi and Cole [14, 15] prepared MgO-doped $Ba_{0.5}Sr_{0.5}TiO_3$ (BST) thin films using metalorganic solution deposition technique. They found that both dielectric loss and insulating characteristics of the doped BST thin films were significantly improved as compared to the undoped one. Other methods, for example, sol-gel process [16], electrophoretic deposition [17] and pulsed laser deposition [18], were also employed to produce MgO-doped BST or MgO-BST composite films. The reduced dielectric losses as a result of the addition of other oxides were at expense of the reduction in dielectric constant, as well as dielectric tunability. A possible explanation to this phenomenon is that Mg substitution into BST shifted the cubic-tetragonal phase transition peak ($T_C$) to a lower temperature, resulting in a decreased dielectric constant at room temperature. The mixing effect suppressed and broaden phase transition peak, which also led to a lower dielectric constant. Both effects were also responsible for the decreased dielectric tunability and improved dielectric loss characteristics [14-16]. However, less information of whether it is possible to use complex oxides to replace simple oxides like MgO.

$LaAlO_3$ is an important material widely used to fabricate microwave dielectric ceramic devices, due to its low dielectric loss and relatively high dielectric constant [19-21]. It possesses a perovskite structure which is similar to BST. These properties of $LaAlO_3$ make it a possible candidate as an additive to improve the dielectric characteristic of BST thin films for microwave tunable applications. Based on this idea, we proposed to deposit LAO-BST thin films. In this paper, deposition and characterization of LAO-BST thin films, on Pt/Ti/SiO$_2$/Si and (100) $LaAlO_3$ substrates, via a pulsed laser deposition (PLD), will be presented. The LAO-BST films were derived from the targets formed by attaching LAO pieces on surface of a BST pellet, which was different from those used to form MgO-doped BST reported by other researchers [18]. The advantage of the combined target configuration was that the film compositions could be easily controlled by the area ratio of LAO and BST. The BST target could also be used to deposit films with other dopants [22]. Our results showed that the LAO-BST thin films were solid solution of LAO and BST and improved dielectric properties were achieved.

## II. Experimental procedure

### A. Target preparation

$Ba_{0.5}Sr_{0.5}TiO_3$ (BST) and $LaAlO_3$ (LAO) targets with a diameter of about 2.5 cm, were prepared using commercial powders of $BaTiO_3$, $SrTiO_3$, $La_2O_3$ and $Al_2O_3$, via the conventional ceramic processing. For $Ba_{0.5}Sr_{0.5}TiO_3$, $BaTiO_3$ and $SrTiO_3$ powders with a ratio of 1:1 were mixed and calcined at 950°C for 1 h before they were compacted and sintered at 1350°C for 4 h, while $LaAlO_3$ target was obtained from the mixture of $La_2O_3$ and $Al_2O_3$, by calcining at 1300°C for 2 h and then sintering at 1550°C for 4 h.

### B. Film deposition

LAO-BST thin films were deposited on Pt/Ti/SiO$_2$/Si and (100) $LaAlO_3$ single crystal substrate, via a pulsed laser deposition (PLD) with a KrF excimer laser at 5 Hz repetition frequency, with an energy of 250 mJ/pulse. The deposition was conducted for 45 minutes, at a substrate temperature of 650°C and a chamber oxygen pressure of 0.2 mbar. The distance between substrate and target was 4.5 cm. A BST target, with a piece of LAO on its surface, was used to deposit LAO-BST thin films. The films with different LAO content, were deposited with the combined targets with different area ratio of LAO. Pure BST and LAO films were also deposited as a comparison. The films derived from the targets of pure BST and BST with 1/6, 2/6 and 3/6 LAO were abbreviated as BSTL0, BSTL1, BSTL2 and BSTL3, respectively.

*C. Structural and electrical characterization*

Phase composition and crystallization of the LAO-BST thin film films were characterized by X-ray diffraction (XRD), using a Philips PW 1729 type X-ray diffractometer with Cu $K_\alpha$ radiation. Surface morphology was examined by scanning electron microscopy (SEM), using a JEOL JSM-6340F type field emission scanning electronic microscope. Film thickness was determined from cross-sectional SEM images.

Dielectric properties of the LAO-BST thin films on Pt/Ti/SiO$_2$/Si substrate were measured using a sandwich structure. Top electrode was Au dots with a diameter of about 0.2 mm, deposited via a vacuum evaporation. An HP 4194 LCR meter was used to record capacitance and dielectric loss as a function of bias voltage, at a frequency of 100 kHz. LAO-BST films on LaAlO$_3$ substrate were measured by a non-destructive microstrip split resonator method at ~7.9 GHz, which was based on the original designe of Galt *et al.* [23-25]. The microstrip split resonator, formed by a straight microstrip line with a 30 μm gap in the center, was patterned on a TMM10i microwave substrate. The films were put on the top of the microstrip line, covering the gap. The dielectric constant $\varepsilon$ and loss tangent $tan\delta$ of the film were derived from the odd-mode resonant frequency, $f_0$ and quality factor $Q$, of the microstrip split resonator. For the study of the electric field dependence of the LAO-BST thin films, a maximum dc voltage of 2.4 kV is applied through two electrode pads on the microstrip circuit board with a gap of about 2.6 mm, corresponding to a maximum electric field of ~9.2 kV/cm.

**III. Results and Discussion**

*A. Phase composition*

Fig. 1 shows the XRD patterns of the LAO-BST thin films deposited on Pt/Ti/SiO$_2$/Si substrates. Pure LAO thin film is also included as a comparison. It is found that the doped samples, together with pure BST and LAO, have all a single phase with perovskite structure. Diffraction peaks coming from Pt are clearly observed in the XRD patterns of all samples. Other peaks indicated by arrows are from the substrate, which was confirmed by the XRD pattern of the uncoated substrate (result not shown). It is, therefore, presumably concluded that the deposited LAO-BST films are solid solutions of LAO and BST.

It is also noticed that the diffraction peaks, corresponding to the perovskite phase, in the XRD patterns, shift to higher angles as LAO content increases. This is obviously demonstrated by (111) diffraction, which is easily distinguished in the pattern of BSTL0, overlaps with the peak of Pt in BST1, and shifts to right of Pt (200) for pure LAO. The shifting of diffraction peak to higher angles in the diffraction patterns indicates a decrease in lattice constant of the LAO-BST thin films with increasing LAO content. It further confirms the above assumption that the LAO-BST films are solid solutions, since the lattice constant of LAO (0.3821 nm) is smaller than that of BST (0.3947 nm). The lattice constants, estimated from the XRD patterns of BSTL1, BSTL2 and BSTL3, are 0.3931, 0.3907 and 0.3891 nm, respectively, corresponding to a composition of $(LAO)_x(BST)_{1-x}$ with x=0.143, 0.278, 0.476.

There a slight variation in preferred orientation occurring in the LAO-BST thin films. The pure BST film is of a random orientation, evidenced by the strongest (110) peak. As LAO is incorporated, (110) peak of BSTL1 becomes less dominant, as shown in Fig. 1 (b), while (100) preferred orientation is observed in BSTL2. However, BSTL3 and pure LAO become random orientation again. This preferred orientation is believed to be related to the film composition and growth behavior. But it is unlikely to affect the dielectric properties of the films, because the effect of composition is more pronounced. Therefore, it is not within the scope of this work. Investigation on this aspect is under way and the results will be reported elsewhere.

Fig. 2 shows the XRD patterns of the LAO-BST thin films on the (100) $LaAlO_3$ substrate. It is shown that the LAO-BST films were epitaxially grown on the (100) LAO substrate, because only diffraction peaks of (100) and (200) are observed in the XRD patterns. Similar to those on Pt/Ti/$SiO_2$/Si substrate, the diffraction peaks also shift to higher 2θ angle as the LAO content increases, indicating again a decreased lattice constant with increasing LAO content. The lattice constants of the LAO-BST thin films on the $LaAlO_3$ substrate are comparable to those of the films on Pt/Ti/$SiO_2$/Si.

*B. Microstructure and surface morphology*

SEM images of the LAO-BST thin films on Pt/Ti/SiO$_2$/Si are shown in Fig. 3. The undoped BST exhibits a uniform surface morphology, with an average grain size of 0.2-0.3 μm. A much dense morphology, with a slightly reduced grain size, is observed in BSTL1. As for high content of LAO, the samples demonstrate a different surface morphology. Some needlelike particles are shown in BSTL2, while clusters of grains appear in BSTL3. It is evidenced that the grain size greatly decreases as the LAO content increases. Fig. 4 shows the cross-sectional SEM images of the LAO-BST thin films. The films thickness of the LAO-BST thin films is about 0.6-0.7 μm. Dense microstructure is observed in BSTL0 and BSTL1, while BSTL2 and BSTL3 are characterized by a columnar structure. This observation is in agreement with the surface images shown in Fig. 3. The microstructure and surface morphology of pure LAO film on Pt/Ti/SiO$_2$/Si were also examined. The results are shown in Fig. 5. The grain size of the LAO film is much smaller than that of BST and LAO-BST films shown in Fig. 3. The LAO film has a thickness similar to that of the pure BST and LAO-BST ones, meaning that LAO and BST have a comparable deposition rate in the present study.

The microstructure of the LAO-BST thin films on the LaAlO$_3$ substrate is much different from that of those on the Pt/Ti/SiO$_2$/Si. Fig. 6 shows the surface image of BSTL1 on LaAlO$_3$. The film consists of uniformly distributed cubic grains, with an average grain size of 70-90 nm. Surface microstructure of LAO on LaAlO$_3$ substrate is shown in Fig. 7, as a comparison.

*C. Electrical properties*

Fig. 8 shows the dielectric constant of the LAO-BST thin films deposited on Pt/Ti/SiO$_2$/Si substrate, measured at 100 kHz, as a function of electrical field. The zero field dielectric constants are 772, 514, 395 and 282, for BSTL0, BSTL1, BSTL2 and BSTL3, respectively. The dielectric constant decreases as a result of the incorporation of LAO into BST due to the low dielectric constant of LAO. The maximum dielectric tunability of the thin films, at the maximum electric field (200 kV/cm), are 65%, 53%, 42% and 14%, respectively. The zero field dielectric losses of the four samples are 0.096, 0.024, 0.023 and 0.027, respectively, which were slightly decrease with applied field. The dielectric properties of our pure BST thin films on Pt/Ti/SiO$_2$/Si substrate are comparable with the literature [26,] [27]. Both dielectric constant and dielectric tunability of the LAO-BST films decrease with increasing LAO content, which is similar to that of BST doped with MgO [12-17].

The effect of LAO on the microstructure and dielectric properties of the BST thin films in the present may not be explained by the mechanism observed in the case of MgO doping, where a combined effect of MgO substitution into BST phase and MgO mixing was proposed [14-16]. That is, the Mg substitution into BST shifted the cubic-tetragonal phase transition peak ($T_C$) to a lower temperature, resulting in a decreased dielectric constant at room temperature. The mixing effect suppressed and broaden phase transition peak, which also led to a lower dielectric constant. Both effects were also responsible for the decreased dielectric tunability and improved dielectric loss characteristics. In the present study, nonferroelectric phase LAO and ferroelectric SBT formed a solid solution. Therefore, the electrical properties of BST might be considered to be "diluted" by LAO.

The variations in dielectric constant of the LAO-BST thin films on $LaAlO_3$ substrates are plotted in Fig. 9. The zero electric field dielectric constants of BSTL0, BSTL1, BSTL2 and BSTL3 are 2436, 2148, 2018 and 1805, while the maximum dielectric tunability of 11%, 13%, 10% and 8%, respectively. The dielectric losses of the undoped and doped samples are all around 0.01, irrespective to the film composition. The high dielectric constant of the films on $LaAlO_3$ substrates, compared to those on Pt/Ti/SiO$_2$/Si, are due to their expitaxial characteristics. The dielectric parameters of the pure BST films are comparable with the values reported in the literature [28]. The reason of the decreased dielectric constant and dielectric tunability for the LAO-BST films on LAO substrate should be the same as that for those films on Pt/Ti/SiO$_2$/Si. The slight increase in dielectric tunability of BSTL1 compared to that of BSTL0 may be attributed to the less mismatch between the film and the substrate in BSTL1. For example, it was reported that an improved dielectric tunability from 36% to 52% for a 57 kV/cm electric field in $Ba_{0.4}Sr_{0.6}TiO_3$ film on $LaAlO_3$ substrate was achieved as a result of the reduction of residual stresses due to thermal annealing [28, 29]. However, such effect was suppressed by the increased content of the nonferroelectric LAO.

## IV. Conclusions

Solid solution LAO-BST thin films were deposited on Pt/Ti/SiO$_2$/Si and (100) $LaAlO_3$ single-crystal substrates, via a pulsed laser deposition. The films on Pt/Ti/SiO$_2$/Si were polycrystalline, with a reduced dielectric constant and tunability and an improved dielectric loss characteristic, while those on LaAlO$_3$ were expitaxial. The compositions of the thin films on both Pt/Ti/SiO$_2$/Si and LaAlO$_3$ substrates were close to the nominal composition of the LAO-BST targets (1/6, 2/6 and 3/6). The dielectric constants of the film derived from pure BST target and those from the targets with 1/6, 2/6 and 3/6 LAO, on Pt/Ti/SiO$_2$/Si substrate, were 772, 514, 395 and 282, with a dielectric loss of 0.096, 0.023, 0.024 and 0.025, and a dielectric tunability of 65%, 53%, 43% and 14%, respectively. The dielectric constant of the corresponding films on LaAlO$_3$ substrate were 2436, 2148, 2018, 1805, with a maximum dielectric tunability of 11, 13, 10 and 8 at a maximum applied voltage of 2.4 kV (~9.2 kV/cm). The result showed that incorporation of LAO into BST is a promising way to improve dielectric properties of BST for the purpose of microwave applications.

**References:**

1. C. H. Mueller, R. R. Romanofsky and F. A. Miranda, IEEE Potentials **20**, 36 (2001).
2. S. S. Gevorgian and E. L. Kollberg, IEEE Trans. Microwave Theory Tech. **49**, 217 (2001).
3. D. S. Korn and H. D. Wu, Integr. Ferroelectri. **24**, 215 (1999).
4. S. Gevorgian, E. Carlsson, E. Wikborg and E. Kollberg, Integrated Ferroelect. **22**, 765 (1998).
5. F. Deflaviis, N. G. Alexepolous and O. M. Staffsudd, IEEE Trans. Microwave Theory Tech. **45**, 963 (1997).
6. S. W. Kirchoefer, J. M. Pond, A. C. Carter, W. Chang, K. K. Agarwal, J. C. Horwitz and D. B. Chrisey, Microwave Opt. Tech. Lett. **18**, 168 (1998).
7. J. B. L. Rao, D. P. Patel and L. C. Sengupta, Integrated Ferroelect. **22**, 827 (1998).
8. C. L. Chen, H. H. Feng, Z. Zhang, A. Brazdeikis, Z. J. Huang, W. K. Chu, F. A. Miranda, F. W. Van Keuls, R. R. Romanofsky and Y. Liou, Appl. Phys. Lett. **75**, 412 (2001).
9. V. Craciun and R. K. Singh, Appl. Phys. Lett., **76**, 1932 (2000).
10. Y. Ding, J. Wu, Z. Meng, H. L. Chan and Z. L. Choy, Mater. Chem. Phys. **75**, 220 (2002).
11. S. S. Gevorgian and E. L. Kollberg, IEEE Trans. Microwave Theory Tech. **49**, 2117 (12001).
12. L. C. Sengupta and S. Sengupta, IEEE Trans. Ultrason. Ferroelect. Freq. Control. **44**, 792 (1997).
13. L. C. Sengupta and S. Sengupta, Mater. Res. Innovat. **2**, 278 (1999).
14. P. C. Joshi and M. W. Cole, Appl. Phys. Lett., **77**, 289 (2000).
15. M. W. Cole, P. C. Joshi, M. H. Ervin, M. C. Wood and R. L. Pfeffer, Thin Solid Films **374**, 34 (2000).
16. E. Ngo, P. C. Joshi, M. W. Cole and C. W. Hubbard, Appl. Phys. Lett. **79**, 248 (2001).
17. M. Jain, S. B. Majumder, R. S. Katiyar, D. C. Agrawal and A. S. Bhalla, Appl. Phys. Lett. 81, 3212 (2002).
18. W. Chang and L. Sengupta, J. Appl. Phys. **92**, 3941 (2002).
19. C. S. Hsu and C. L. Huang, Mater. Res. Bull. **36**, 1939 (2001).
20. D. G. Lim, B. H. Kim, T. G. Kim and H. J. Jung, Mater. Res. Bull. **34**, 1577 (1999).
21. C. L. Huang, R. J. Lin and H. L. Chen, Mater. Res. Bull. **37**, 449 (2002).
22. G. Yang, W. Wang, Y. Zhou, H. Lu, G. Yang and Z. Chen, Appl. Phys. Lett. **81**, 3969 (2002).

23. D. Galt, J. C. Price, J. A. Beall and R. H. Ono, Appl. Phys. Lett. **63**, 3078 (1993).

24. D. Galt, J. C. Price, J. A. Beall and T. E. Harvey, IEEE Trans. Appl. Supercond. **5**, 2575 (1995).

25. C. Y. Tan, L. F. Chen, K. B. Chong, Y. T. Ngow, Y. N. Tan and C. K. Ong., Rev. Sci. Instrum. (submitted).

26. Y. A. Jeon, W. C. Shin, T. S. Seo and S. G. Yoon, J. Mater. Res. **17**, 2831 (2002).

27. C. Basceri, S. K. Streiffer, A. I. Kingon and R. Waser, J. Appl. Phys. **82**, 2497 (1997).

28. C. M. Carlson, T. V. Rivkin, P. A. Parilla, J. D. Perkins, D. Ginley, A. B. Kozyrev, V. N. Oshadchy and A. S. Pavlov, Appl. Phys. Lett. **76**, 1920 (2000).

29. Z. G. Ban and S. P. Alpay, J. Appl. Phys. **93**, 504 (2003).

**Figure Captions**

FIG. 1 XRD patterns of the LAO-BST thin films on Pt/Ti/SiO$_2$/Si substrate: (a) BSTL0, (b) BSTL1, (c) BSTL2, (d) BSTL3 and (e) LAO.

FIG. 2 XRD patterns of the LAO-BST thin films on LaAlO$_3$ substrate: (a) BSTL0, (b) BSTL1, (c) BSTL2 and (d) BSTL3 (F: film and S: substrate).

FIG. 3 Surface SEM images the LAO-BST thin films on Pt/Ti/SiO$_2$/Si substrate: (a) BSTL0, (b) BSTL1, (c) BSTL2 and (d) BSTL3.

FIG. 4 Cross-section SEM images the LAO-BST thin films on Pt/Ti/SiO$_2$/Si substrate: (a) BSTL0, (b) BSTL1, (c) BSTL2 and (d) BSTL3.

FIG. 5 Surface and cross-section SEM images of LAO deposited on the LaAlO$_3$ substrate.

FIG. 6 SEM image of BSTL1 thin film on LaAlO$_3$ substrate.

FIG. 7 SEM image of LAO film on LaAlO$_3$ substrate.

FIG. 8 Dielectric constant of the LAO-BST thin films on Pt/Ti/SiO$_2$/Si substrate (measured at 100 kHz) as a function of electrical field.

FIG. 9 Dielectric constant of the LAO-BST thin films on LaAlO$_3$ substrate (measured at ~6.8 GHz) as a function of applied voltage.

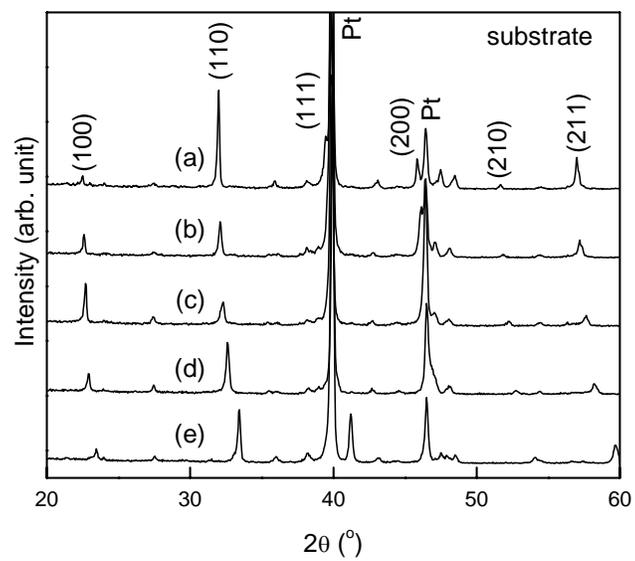

Kong *et. al.*,    Fig. 1/9

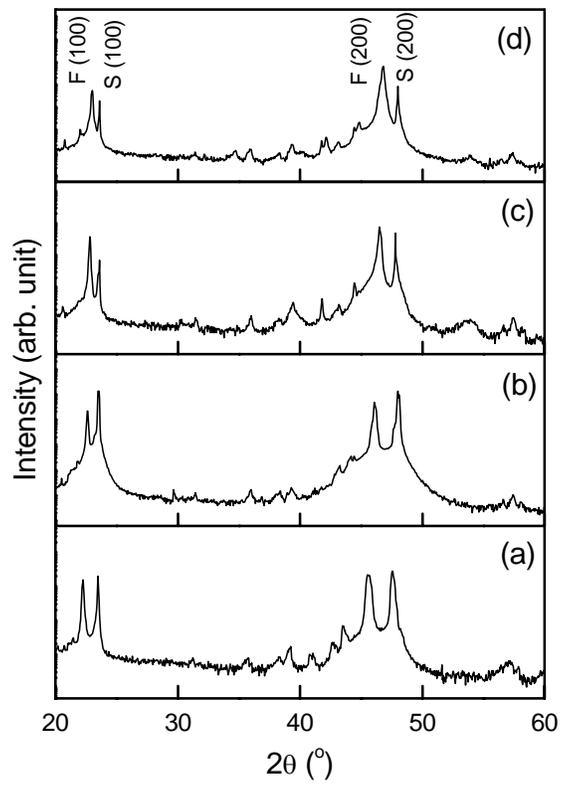

Kong *et. al.*, Fig. 2/9

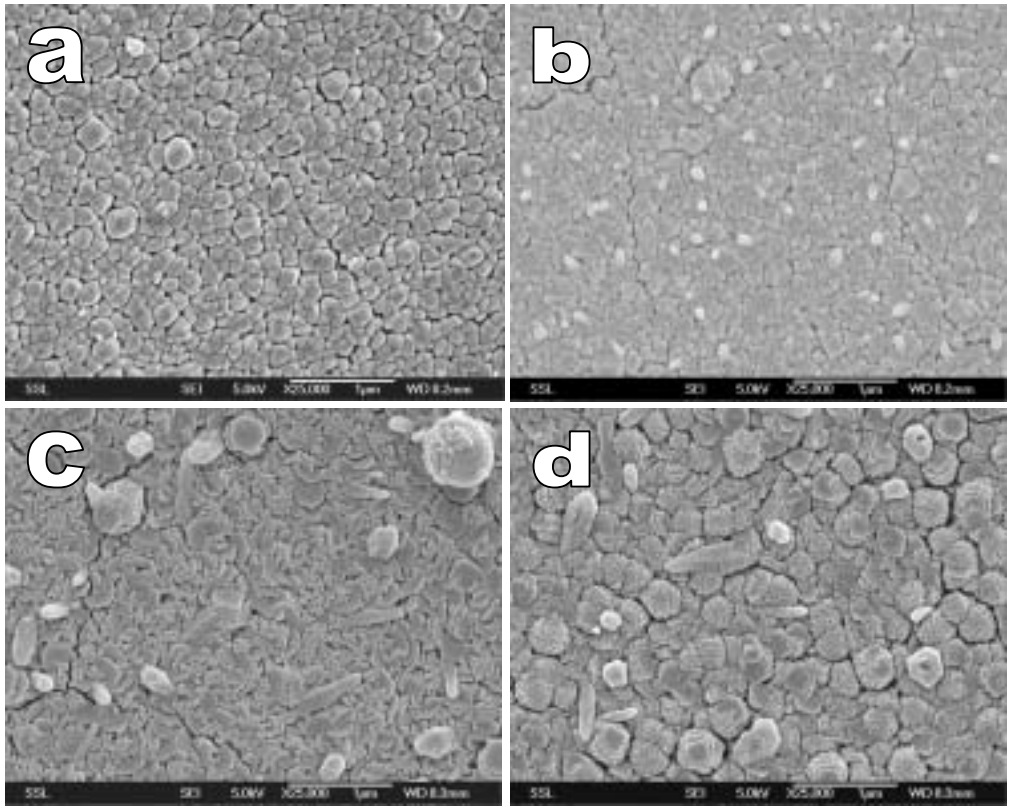

Kong *et. al.*,   Fig. 3/9

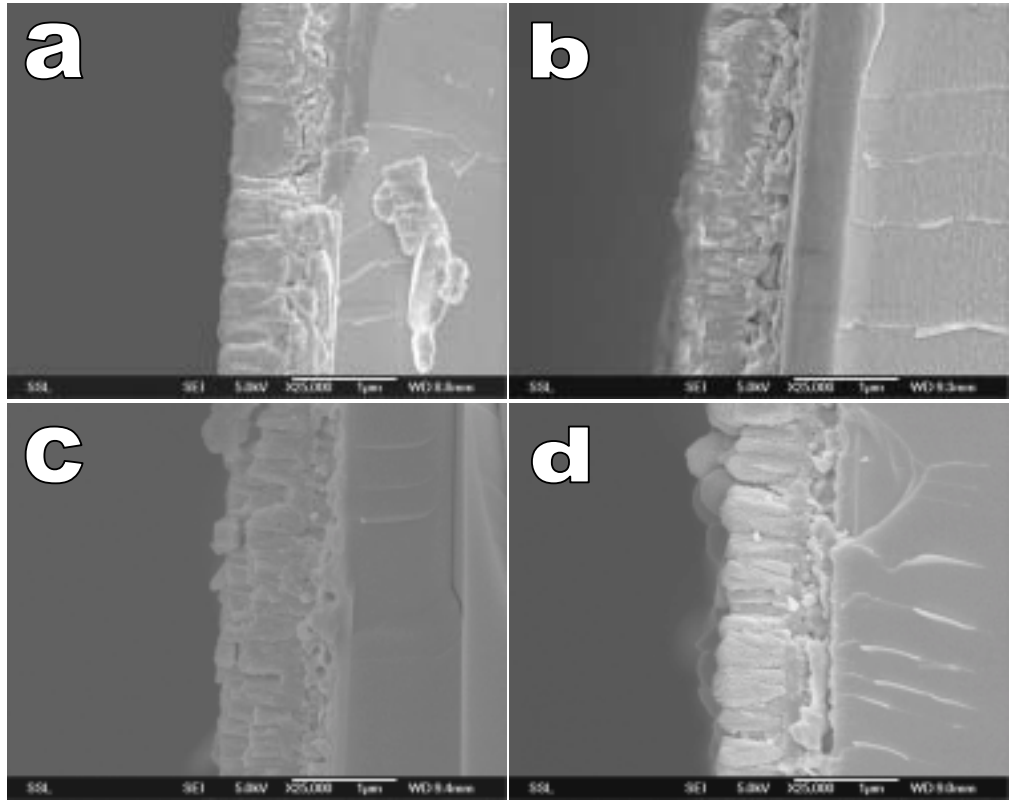

Kong *et. al.*,   Fig. 4/9

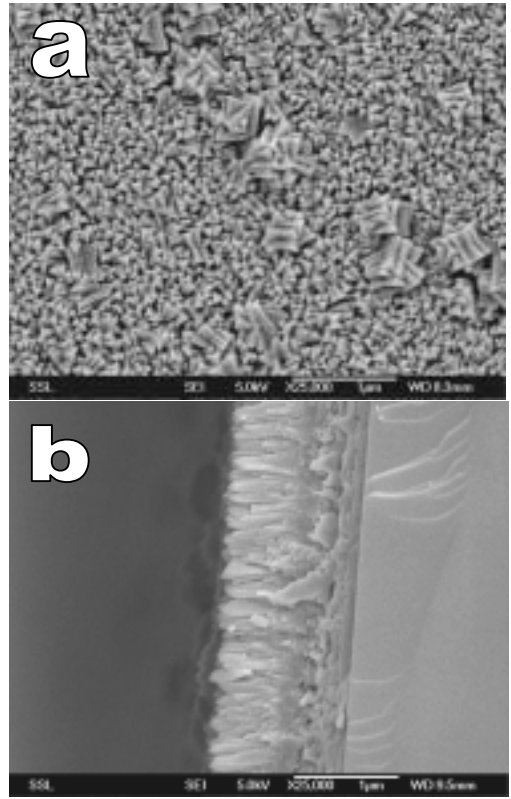

Kong *et. al.*,   Fig. 5/9

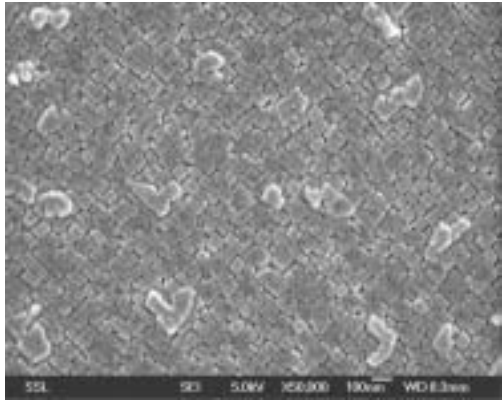

Kong *et. al.*,    Fig. 6/9

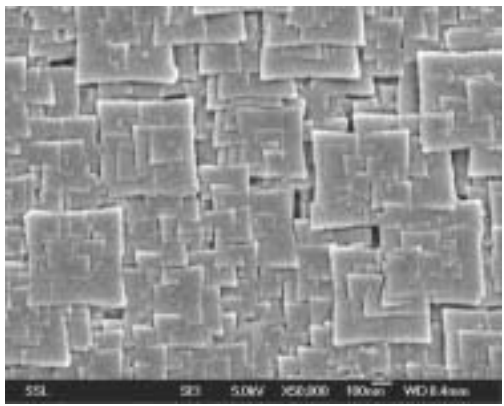

Kong *et. al.*,    Fig. 7/9

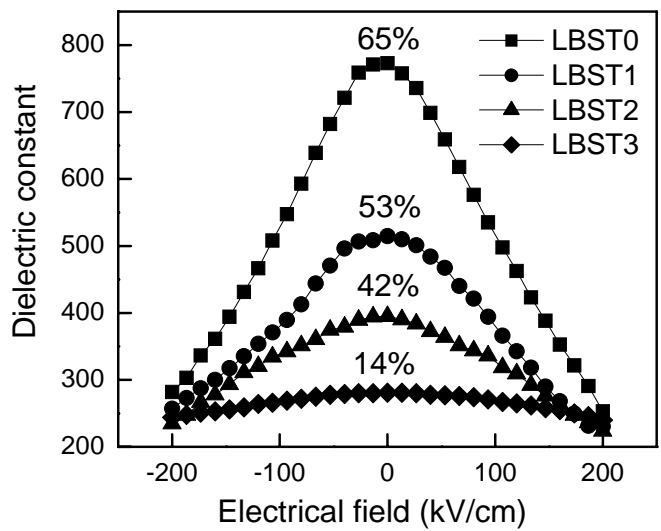

Kong *et. al.*,  Fig. 8/9

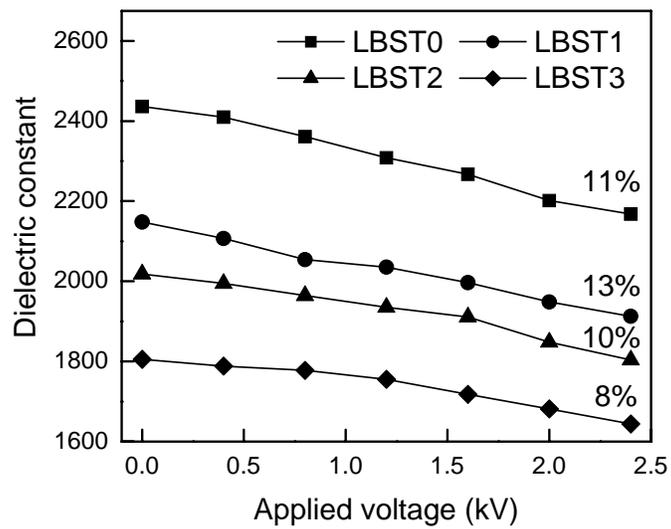

Kong *et. al.*, Fig. 9/9